\documentclass{article}
\usepackage[utf8]{inputenc}
\usepackage[hyphens]{url}
\usepackage{authblk}
\usepackage{multirow}
\usepackage{booktabs}
\usepackage[table]{xcolor}
\usepackage{graphicx}
\providecommand{\keywords}[1]{\textbf{\textit{Keywords:}} #1}

\author[1,2]{Constantinos Patsakis}
\author[1]{Fran Casino}
\author[1]{Nikolaos Lykousas}
\author[3]{Vasilios Katos}
\affil[1]{University of Piraeus}
\affil[2]{Athena Research Center}
\affil[3]{Bournemouth University}
\title{Unravelling Ariadne's Thread: Exploring the Threats of Decentalised DNS}
\date{}
\begin{document}

\maketitle

\begin{abstract}
The current landscape of the core Internet technologies shows considerable centralisation with the big tech companies controlling the vast majority of traffic and services. This has sparked a wide range of decentralisation initiatives with perhaps the most profound and successful being the blockchain technology. In the past years, a core Internet infrastructure, domain name system (DNS), is being revised mainly due to its inherent security and privacy issues.
One of the proposed panaceas is Blockchain-based DNS, which claims to solve many issues of traditional DNS. However, this does not come without security concerns and issues, as any introduction and adoption of a new technology does - let alone a disruptive one such as blockchain. In this work, we discuss a number of associated threats, including emerging ones, and we validate many of them with real-world data. In this regard, we explore a part of the blockchain DNS ecosystem in terms of the browser extensions using such technologies, the chain itself (Namecoin and Emercoin), the domains, and users which have been registered in both platforms. Finally, we provide some countermeasures to address the identified threats, and we propose a fertile common ground for further research.
\end{abstract}


\section{Introduction}
One could argue that there is a periodic paradigm shift between centralisation and decentralisation in computer science. A representative example is the transition from mainframes with dummy terminals to personal computers or the shift from centralised local storage to the cloud. Although the Internet was in principle designed to be distributed and decentralised by nature, in reality, the control is placed onto a relatively limited number of stakeholders and the quest for further decentralisation is becoming an imminent need. Such requirement manifests in many ways, see for example the case of net neutrality, or the concept of crowdsourcing which also attempts to address efficiency and sustainability issues. As such, in recent years, we are witnessing an increasing demand and creation of decentralised services.

The most profound example is the blockchain technology, which is being widely deployed in various and different fields \cite{casino2018systematic}. In different forms, this decentralisation shift is gradually being realised in traditionally centralised services, such as DNS. DNS is a distributed database with a centralised data governance model, primarily controlled by ICANN. In this regard, ICANN manages the top-level domains (TLDs), and the operation of root name servers. In practice, once a client needs to contact a server for which it knows the name but not its IP address, it performs a query to a particular DNS server which is known through some network configuration protocol. Depending on how often this domain is requested, it may be hosted in the cache of the DNS server. If this is not the case, the query is propagated to the root name server which can find the servers for the corresponding TLD and then forward the query to the corresponding authoritative name server which would return the corresponding IP.

While DNS is currently one of the oldest still working Internet application-level protocols, it has several drawbacks that mandate its replacement. For instance, DNS does not support cryptographic primitives. Therefore, any query and response can be intercepted by anyone in the same network, implying many privacy issues. Moreover, anyone on the same network may inject a response of an intercepted query, indicating many security risks. Furthermore, some regimes have exploited DNS to censor unwanted web pages and services, and in the past years, DNS servers have been used in amplification denial of service attacks and their records have been poisoned.

\noindent{\bf Motivation:} All the aforementioned issues have driven many researchers to seek for alternative solutions to DNS. This is evident for example with the shift towards, e.g. DNS over HTTPS\footnote{\url{https://tools.ietf.org/html/rfc8484}} and DNS over TLS\footnote{https://tools.ietf.org/html/rfc7858} and 8310\footnote{https://tools.ietf.org/html/rfc8310}, while others are shifting looking for solutions beyond ICANN support. However, these solutions are still centralised. One of the most promising decentralised solutions is blockchain DNS which is already adopted by several chains, e.g. Ethereum, Namecoin, and Emercoin, or specific protocols. In fact, despite their infancy, blockchain domains have attracted the interest of several big players. A notable example is Alibaba which recently filed a patent for a blockchain-based management domain name management system\footnote{\url{https://domainnamewire.com/wp-content/alibaba-blockchain-domain-patent.pdf}}. A brief overview of blockchain DNS and some scepticism was initially provided in \cite{8515149}. So far, the proliferation of blockchain DNS projects and research proposals (as later discussed in Section \ref{sec:related}), are already being exploited by cybercriminals\footnote{\url{https://www.digitalshadows.com/blog-and-research/how-cybercriminals-are-using-blockchain-dns-from-the-market-to-the-bazar/}}. Therefore, we argue that there is a need for exploring threat models relating to novel blockchain solutions \footnote{\url{https://en.bitcoinwiki.org/wiki/Blockchain_Projects_List}}, as well as decentralised file storage systems \cite{ijisipfs}. The decentralisation of services may undoubtedly provide many new features in terms of privacy, security and democratisation; nevertheless, sudden changes in the backbone of well-established services and infrastructures may come at a significant cost. Adversaries are expected to take advantage of such changes, by exploiting the lack of knowledge, experience and maturity of the users, as well as inherent flaws that are present in the early stages of a new technology. At the same time, the use of encrypted and covert communications adds another layer of difficulty to detect infected systems \cite{cose2019}, for instance, in the case of botnets. Therefore, it is imperative to raise awareness about the opportunities as well as the introduced security threats. Moreover, a discussion and exploration of possible countermeasures against the next generations of malware campaigns is also imperative. This paper aims to fill this research gap providing an overview of the current state of the art and practice (Section \ref{sec:related}) and a detailed presentation of the emerging threats and how they could be amplified (Section \ref{sec:threats}). Further to merely speculating future threats, we perform an investigation and analysis of such ecosystem and illustrate that the landscape of converting these threats to actual risks is already present. To this end, in Section \ref{sec:experiments}, we showcase the results of an in-depth analysis of Namecoin, Emercoin and Blockchain DNS crawling. In this regard, note that the threats discussed and the outcomes retrieved form the statistical analysis could be extended to other Blockchain DNS systems. Finally, some remarks and findings are further discussed, along with possible countermeasures in Section \ref{sec:conclusions}. To the best of our knowledge, this is the first paper that provides a discussion about the current state of practice and identification of novel threats beyond the actual state-of-the-art supported by data analysed from real-world blockchain sources.

\section{Background}
\label{sec:related}
\subsection{Blockchain-based DNS}

Blockchain-based DNSs have been receiving an ever increasing attention in recent years \cite{8515149}. According to Scopus, Web of Science and Google Scholar, a set of approaches, some of which are fully functional, have emerged in the literature since 2016. In what follows, we describe them and analyse the main features of the most relevant and adopted solutions.

The work presented by Hari et al. \cite{Hari2016204} is one of the first works that propose the use of blockchain to develop a DNS infrastructure. The authors discuss the benefits of such a system over the main threats and drawbacks of traditional models such as compromised hosts, spoofing, trust management, and its heavy dependence on PKIs.
\cite{Benshoof20161279} proposed a system named D\textsuperscript{3}NS, which uses a distributed hash table and a domain name ownership implementation based on the Bitcoin blockchain. They aim to replace the top-level DNS and certificate authorities, offering increased scalability, security and robustness. \cite{Liu2018189} proposed a blockchain-based decentralisation DNS resolution method with distributed data storage to mitigate single points of failure and domain name resolution data tampering. \cite{Gourley2019173} proposed the use of blockchain to enhance the certificate validation procedure to create an improved DNS security extension, providing the same benefits with DNSSec while overcoming its main drawbacks. Similarly, in an attempt to reduce the level of trust in certificate authorities, \cite{Guan2019345} presented AuthLedger, a blockchain-based system that provides efficient and secure domain name authentication. \cite{1007978} introduced BlockZone which uses a replicated network of nodes to offer efficient name resolution through an improved PBFT consensus mechanism.

Some work focused on IoT systems and their communication protocols has also been proposed. For example, \cite{Duan20181} presented DNSLedger, a hierarchical multi-chain structure in which domain name management and resolution are performed in a decentralised way. The authors claim that their system could enhance IoT-related communication technologies due to its efficiency. BlockONS, proposed by \cite{Yoon2019219}, is a system that aims to overcome classical problems related to DNS resolution, namely DNS cache poisoning, spoofing, and local DNS cracking. Authors propose a robust and scalable object name service, especially relevant for the IoT ecosystem. ConsortiumDNS was introduced by \cite{Wang2018617} as a system based on a three-layer architecture composed by consortium blockchain, a consensus mechanism and external storage. Authors claim that their approach increases the efficiency of the overall system, compared to other well-known approaches such as Namecoin or Blockstack. Finally, we may also find some patented designs of Blockchain-based DNS systems in \cite{li2019blockchain,li2019systems}.

Currently, there are several relevant and widely adopted blockchain-DNS projects. Handshake \footnote{\url{https://handshake.org/}} is one of the most widely supported technologies, which aims at creating an alternative to existing certificate authorities. Therefore, Handshake aims to replace the root zone file and the DNS name resolution and registration services worldwide. The Ethereum name service (ENS)\footnote{\url{https://ens.domains/}} uses smart contracts to manage the \texttt{.eth} registrar by means of bids and recently added the support for \texttt{.onion} addresses. Namecoin\footnote{\url{https://www.namecoin.org/}} is a cryptocurrency, based on Bitcoin, with additional features such as decentralised name system management, mainly of the \texttt{.bit} domain. It was the first project to provide a solution to Zooko's triangle since their system is secure, decentralised and human-meaningful. Nevertheless, contrary to well-established blockchains like Bitcoin, Namecoin's main drawback is its insufficient computing power, which makes it more vulnerable to the 51\% attack. Practically, if an adversary manages to get a slight majority of the computing power, she may rewrite the whole chain. Blockstack \cite{196208} is a well-known blockchain-based naming and storage system that overcomes the main drawbacks of Namecoin. Blockstack's architecture separates control and data planes, enabling seamless integration with the underlying blockchain. EmerDNS\footnote{\url{https://emercoin.com/en/documentation/blockchain-services/emerdns/emerdns-introduction}} is a system for decentralised domain names supporting a full range of DNS records. EmerDNS operates under the ``DNS'' service abbreviation in the Emercoin NVS. Nebulis\footnote{\url{https://www.nebulis.io/}}  is a globally distributed directory that relies on the Ethereum ecosystem and smart contracts to store, update, and resolve domain records. Moreover, Nebulis proposes the use of off-chain storage (i.e. IPFS) as a replacement for HTTP. OpenNIC\footnote{\url{https://www.opennic.org/}} deserves a special mention since it is a hybrid approach in which a set of peers manages namespace registration, yet the name resolving task is fully decentralised. OpenNIC provides DNS namespace, and resolution of a set of domains, some of them agreed with Blockchain solutions such as EmerDNS and New Nations\footnote{http://www.new-nations.net/}, the latter being a TLD provider for nation-states that have not received a country code top-level domain (ccTLD). Moreover, OpenNIC resolvers have recently added access to domains administered by ICANN. In addition to namespace registrar, users can also create their own TLD on request. Table \ref{tab:features} summarises the main features of the most relevant Blockchain-DNS systems.


 \begin{table}[!ht]
   \rowcolors{2}{gray!25}{white}
   \setlength{\tabcolsep}{6pt}
   \scriptsize
   \caption{Technical characteristics of the most relevant DNS systems. Although Blockstack is blockchain agnostic, it is mainly used with Bitcoin blockchain.}
  \begin{tabular}{llp{1in}p{1in}}
     \toprule
   \textbf{Method} &  \textbf{Pedigree Platform} &   \textbf{Registrar and Resolution Management}     &\textbf{TLD Examples}  \\
   \midrule
 ICANN & Network of Servers and resolvers &  Centralised  & \texttt{.com $\vert$ .net $\vert$ .org}  \\
  OpenNIC & Decentralised Servers & Hybrid  & \texttt{.bbs $\vert$ .pirate $\vert$ .libre}  \\
  ENS &  Ethereum &  Decentralised  & \texttt{.eth $\vert$ .onion}  \\
    Handshake & Bitcoin &  Decentralised  & unrestricted  \\
   Blockstack & Blockchain agnostic  &    Decentralised   & \texttt{.id $\vert$ .podcast $\vert$ .helloworld} \\
   Emercoin &  Bitcoin   &   Decentralised    &  \texttt{.coin $\vert$ .bazar $\vert$ .emc} \\
     Namecoin &   Bitcoin and Peercoin  &   Decentralised    & \texttt{.bit} \\
     \bottomrule
   \end{tabular}
   \label{tab:features}
 \end{table}

\begin{figure}[ht]
    \centering
    \includegraphics[width=.8\textwidth]{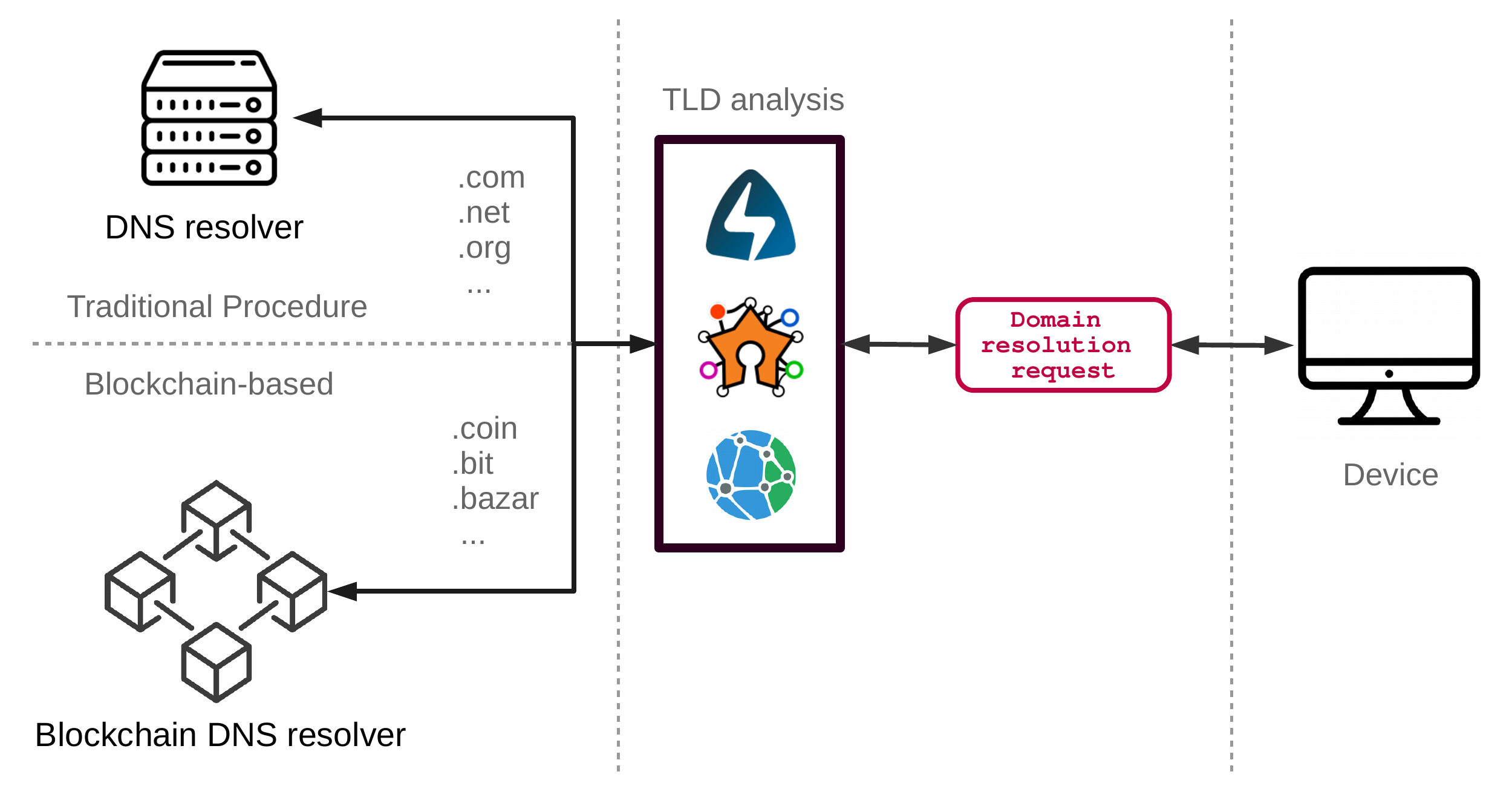}
    \caption{Workflow of the browser extensions procedure to enable resolution of EmerDNS, Namecoin, New Nations and OpenNIC domains. The extension analyses the TLD extension of the requested domain and directs the query to the corresponding DNS system.}
    \label{fig:bdnsextension}
\end{figure}

Internet users can reach the TLDs offered by Namecoin, OpenNIC, New Nations, and EmerDNS (e.g. \texttt{.bit}, \texttt{.coin}, \texttt{.emc}, \texttt{.lib} and \texttt{.bazar}) through various browser extensions such as \texttt{peername}, \texttt{blockchain-DNS} and \texttt{friGate} \cite{Emercoinresources}. Their modus operandi is described in Figure \ref{fig:bdnsextension}.

\subsection{Distributed platforms and C2}

Nowadays, sophisticated malware campaigns are continuously emerging, focusing on exploiting decentralised technologies such as blockchain and distributed file storage (DFS). In the case of botnets, the use of technologies such as DFS systems prevents the generation of NXDomain responses, which is a well-known indicator of compromise type for malware using domain generation algorithms. In this regard, \cite{ijisipfs} extended the definition of domain generation algorithms (i.e. a family of pseudo-random domain name generators to which an infected host will try to connect to find the C2 server) into a more generic framework, namely Resource Identifier Generation Algorithms (RIGA). Moreover, the authors showed how DFS like IPFS could enhance malware campaigns due to their attractive properties, such as immutability, efficiency and negligible costs. Botnet C2 management through Blockchain systems is also a threat, as proposed in \cite{Ali2018} and used in the case of the Cerber ransomware, analysed in \cite{Pletinckx2018}. In this case, the malware finds the C2  based on transaction information of the bitcoin blockchain. A more recent threat is the use of encrypted and covert communication channels such as in the case of DNSSec, DNS over HTTPS (DoH) and DNS over TLS (DoT). Although these technologies, like in the previous cases, hinder the possibility of using NXDomain information leaks to detect suspicious behaviour, Patsakis et al. showed that even in such case some patterns might emerge \cite{cose2019}, which can be used to identify and classify DGA families accurately. Regarding recent Blockchain-DNS systems, there exist a set of attack vectors that can be exploited, such as in recent cases\footnote{\url{https://www.digitalshadows.com/blog-and-research/how-cybercriminals-are-using-blockchain-dns-from-the-market-to-the-bazar/}}.

\section{The decentralised DNS threat}
\label{sec:threats}
As previously stated, blockchain-based DNSs provide a set of characteristics, which are summarised in Table \ref{tab:benefitsBDNS}. In this regard, one could argue that the traditional DNS seems to be outdated, compared to the novel Blockchain DNSs. Nevertheless, traditional DNS proofed their reliability and scalability from early 80's until today with modest adjustments. Moreover, blockchain-based DNSs exhibit a set of potential threats and attack vectors that must be considered\cite{amado18,sanders19,ali2019blockchain,rasmussen2013surveying}. In the following sections, we analyse the most well-known threats and identify novel ones as well as discuss their possible impact for the system and the final users.

\begin{figure}
    \centering
    \includegraphics[width=.8\textwidth]{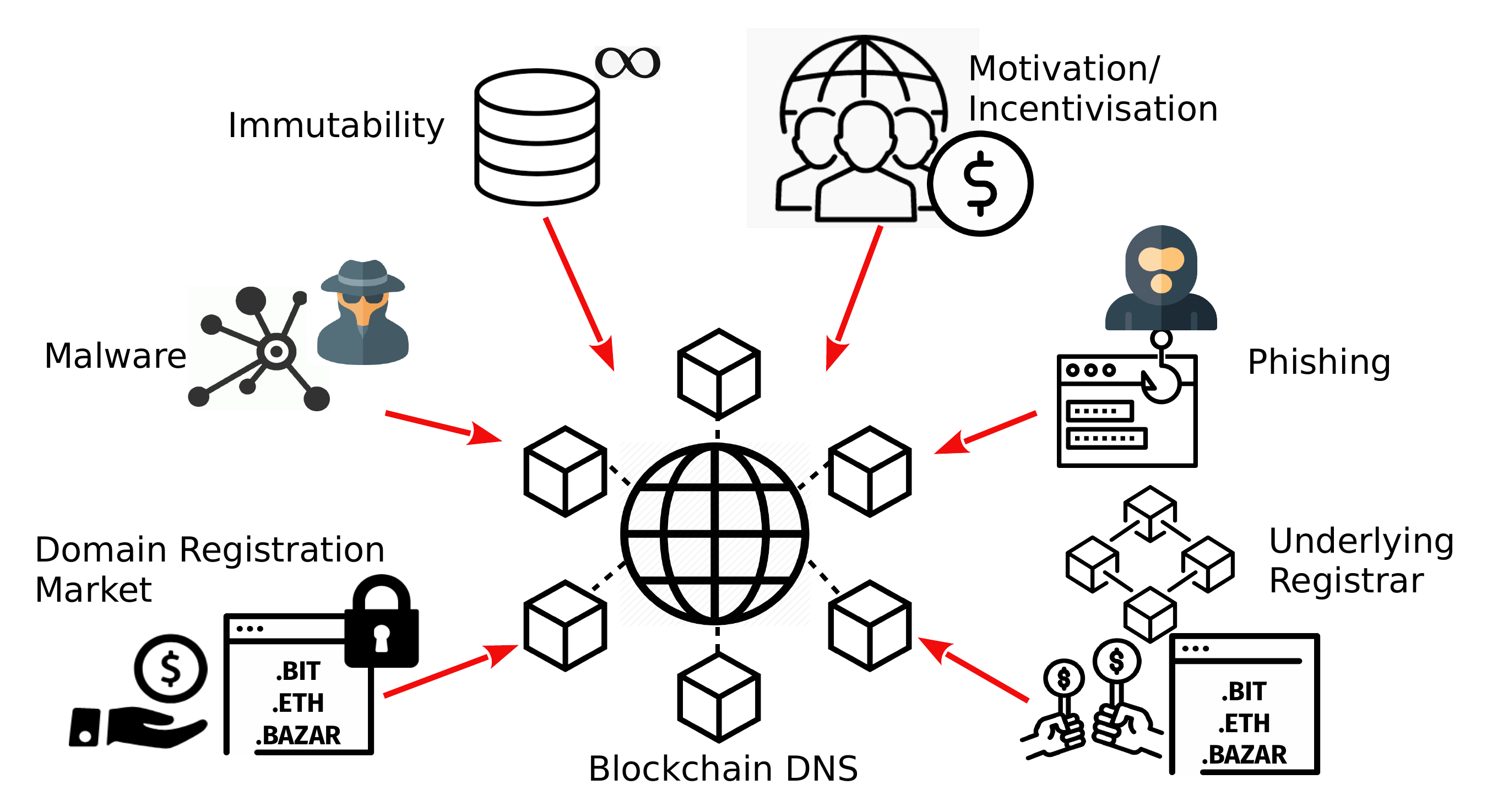}
    \caption{An overview of main threats of blockchain DNSs.}
    \label{fig:mainthreats}
\end{figure}

A summary of the emerging threats due to the adoption of blockchain DNSs is depicted in Figure \ref{fig:mainthreats}.

\begin{table}[ht]
 \rowcolors{2}{gray!25}{white}
   \scriptsize
   \caption{Main characteristics of blockchain DNSs.}
  \begin{tabular}{p{.25\textwidth}p{.7\textwidth}}
     \toprule
   \textbf{Property} &  \textbf{Description}  \\
   \midrule
Trust       & Verifiable and robust consensus mechanisms \\
      Decentralisation &  The network is totally distributed with no central entities \\
      Availability  & The availability of the network depends on multiple peers and not on a single entity. \\
      Censorship-resistant &Access to information and domain name resolution are not subject to borders or bans \\
 Robustness  &  Resilient to attacks that affect centralised DNS systems such as MiM, spoofing, cache poisoning, cracking. \\
  Unlimited Resources & A high number of simultaneous users sharing their assets. \\
Namespace Freedom & Registration of new SLDs and TLDs \\
Automated Management &  Auctions to register domain names, fast and transparent ownership control \\
     \bottomrule
   \end{tabular}
   \label{tab:benefitsBDNS}
 \end{table}


\subsection{Malware}
\label{sec:threatmalware}
In the case of malware, blockchain-based DNSs offer considerable potential. Employing such technologies unlock the capability to register a substantial number of domains with low entropy, which were not available in the market. Currently, malware authors are using DGAs to generate domain names (i.e. algorithmically generated domains or AGDs); however, since most short and meaningful domain names are not available, they resorted to the use of long and random-looking domain names. Therefore, a compromised host which uses a DGA to resolve the C2 server issues many Non-Existent Domain (NXDomain) requests which can be analysed and the attribution to the proper DGA can be made efficiently.

With the use of blockchain-based DNS systems, the conventional NXDomain requests will not be issued (see next section), hence hindering the detection mechanisms. Moreover, by using domain names with lower entropy, many filtering and machine learning algorithms are rendered useless. The latter practice is exposed in Section \ref{sec:experiments}.

Even more, the use of blockchain-based DNSs implies further issues for malware analysts. When performing static analysis of the malware and its reverse-engineered code, the analyst and the tools that she uses must be aware of the new domains. Traditional filters for domain names will fail, for instance, to reveal calls to \texttt{.bit} domain as the resolution mechanism is completely different. However, requests to traditionally benign domains, e.g. to \url{google.com}, may resolve to a completely different IP and the same applies for case sensitive domains, e.g. to \url{GoOgle.com}, or the use of spaces, e.g. \url{goog}\textvisiblespace\url{le.com}. While Handshake, for instance, may have already taken some precautionary measures for the highly visited domain names, this does not prevent the use of existing domain names with less visibility being exploited to serve malware. Unfortunately, as discussed in the next section, our data indicate that this attack vector is something that will be used in the near future. Finally, it should be noted that an adversary could easily perform fast fluxing and change the IP addresses that are used whenever deemed necessary. As reported by deteque\footnote{\url{https://www.deteque.com/news/abused-top-level-domains-2018/}}, more than 100 domains registered in blockchain DNS registrars were used by C2 servers in 2018 implying that their use is actively being exploited by cybercriminals.


\subsection{Underlying registrar mechanism}
The primary methodology to register domains in \sloppy{blockchain} DNSs is to perform bids or auctions, being the first-request, first-served an outdated strategy. Nevertheless, in the case of a vulnerability in the underlying bid system, users can get control of domains as recently happened\footnote{\url{https://www.coindesk.com/ethereum-name-service-auction-exploited-to-grab-apple-domain-and-it-cant-be-undone}}. Moreover, most blockchain DNS systems such as Emercoin allow the registration of case sensitive domains (as we will discuss in Section \ref{sec:experiments}), something infeasible in traditional systems. The latter, if paired with some other unrestricted practices such as the use of spaces, non UTF-8 or Ascii characters, may end up with an uncontrollable amount of alternative domain names which may not be easily distinguishable from the legitimate ones. Although this may be a target scenario for malicious actors, this situation may have an impact to the trust and the will of users towards the system. Note that these practices could be prevented and reverted in traditional DNS, but not in blockchain DNSs. Therefore, careful design of the methodology, as well as a proper implementation of the underlying smart contracts, should be carried out to prevent this kind of behaviours. In the case of systems that offer the use of DNS name resolving services and the registrar of some TLD, a way to prevent this is blacklisting them, although being a controversial strategy. This threat is critical in systems like Handshake and others that may arise, aiming at a full substitution of traditional mechanisms such as ICANN, since legitimate names that are owned by an organisation in a conventional ICANN-supported DNS may end up being controlled by malicious users. 

In essence, an uncontrolled and fully decentralised DNS type of service may lead to having \textit{parallel} Internets. Note that each blockchain DNS system enables the registration of arbitrary sets of TLDs, which may overlap with existing ones. Therefore, the same domain would resolve to different IPs, depending on the blockchain DNS system used. For instance, even if not used, the domain \url{google.com} is registered in Emercoin in block 252362\footnote{\url{ https://explorer.Emercoin.com/block/252362}}. This opens a whole new scenario of possibilities, in which users can have access to a myriad of contents without restriction, yet in most cases, they could be owned by a malicious entity. The latter problem, as we will discuss in the following paragraphs, is exacerbated by other properties such as immutability.

\subsection{Domain registration market}
In the least sinister scenario, we consider the case of one registering the domain name of an existing, legitimate webpage. Since the blockchain TLDs are not known to the vast majority of people, it is expected that many people will rush to buy such names requesting a good payment in exchange for the name. As discussed in more detail in Section \ref{sec:experiments}, this is not only a hypothesis but a real case. Block 160356 of Emercoin\footnote{\url{https://explorer.emercoin.com/block/160358}} illustrates such requests were the fees range from \$600 to \$20,000.

The problem is actually an extension of domain backordering as in this instance we are not dealing with expired domains, but with new TLDs. The existence of ICANN and intermediates, e.g. registrars, allows in many cases the arbitration or even the shutdown or handing over of a domain name; however, the use of blockchain systems does not allow for such mechanisms to be applied. In fact, at the time of writing, one can register a name for an arbitrary amount of time in Emercoin. For instance, there are many domain names in Emercoin which are registered for thousands of years, e.g. there are domains registered up to 5014 and 12012 in blocks 200590 and 380209, respectively\footnote{\url{https://explorer.emercoin.com/block/200590} and \url{https://explorer.emercoin.com/block/380209}}.

\subsection{Phishing}

Phishing is a fraudulent practice which targets an audience to obtain valuable personal information by using impersonation of entities, persons and more techniques. According to State of the phish 2019 by Proofpoint\cite{proofpoint2019}, the number of compromised accounts by these attacks varied from 38\% to 65\% from
2017 to 2018. This type of cyber-attack leverages socially engineering methods to trick users into performing activities that in some way; most usually monetary, will benefit the attacker \cite{khonji2013phishing}. Email is by far the most widely used method to date to perform phishing is email is the most popular avenue for a phishing attack, with more
than 90\% of successful cyber-attacks/security breaches starting from a spoofed email\cite{phishme2016}. In fact, the automated
nature of this attack, coupled with the incapacity of users to determine a phishing attack \cite{iuga2016baiting} make the threat even more dangerous. There are many factors which augment this threat and most reside on the human-side aspect of the problem. For instance, the timing of the attack, the authoritarian writing, as well as the exploitation of common practices in an organisation, may significantly bias the user into accepting the email as legitimate.
Clearly, the use of spoofed or compromised email accounts further complicates the situation.

In the context of blockchain DNSs, the problem is amplified. The users are accustomed to visiting specific web pages and sending emails to particular accounts. If these accounts are pointing to a similar address, e.g. changing the TLD, many users are for sure expected to be tricked. The use of punycodes for phishing or the use of different TLDs can be considered a norm in phishing. With the introduction of blockchain DNSs, an adversary has far more options as there is a wide range of domains that are becoming available at a minimum cost. 
Practically, this means that not only the phishing sites may have a similar domain name with legitimate ones, but with the use of, e.g. a Let's Encrypt\footnote{\url{https://letsencrypt.org}} certificate, the fraudulent web pages may have valid and trusted HTTPS support. Therefore, the phishing page may have all the distinctive elements, from the UI, the HTTPS support and the valid domain name, making it very difficult for a common user to distinguish the original from the phishing page.

\subsection{Lack of motivation}
Motivation under the blockchain DNS paradigm is clearly related to the features offered by such a system, including censorship resistance as one of the main attractions. Nevertheless, these desirable features come at a cost, since decentralised systems totally rely on their network of nodes and their participation. Therefore, keeping the user's interest in blockchain DNSs is critical.

Unarguably, blockchain's adoption in a myriad of scenarios is a reality \cite{casino2018systematic}. Nevertheless, not all blockchain-based projects succeed. In this regard, according to statistics retrieved from Deadcoins\footnote{https://deadcoins.com/} there exist approximately 1000 dead cryptocurrencies and more than 660 fraudulent cryptocurrency attempts. 
Interestingly enough, as of 2018, ICO scams have already raised more than 1,000 million dollars\footnote{https://www.ccn.com/ico-scams-have-raised-more-than-1-billion-report-claims/}. Despite the existence of some awareness campaigns such as HoweyCoin\footnote{https://www.howeycoins.com/index.html}, the lack of a specific and interoperable framework to pursue such deviant behaviour enables the persistence of these practices. In the case of blockchain, this may hinder the creation of new projects as well as the persistence of well-known and established ones. One of the main problems that could arise is an unbalanced/unstable computational power, which could compromise the underlying consensus mechanisms and trigger, for instance, a 51\% attack. Note that this attack may be applied regardless of the number of users that use a blockchain DNS solution, as the attack is targeted towards the nodes that store the blockchain which, depending on the rewards they have, their participation may decrease over time. The latter may allow an adversary to control the blockchain and compromise its integrity without having to exploit any, e.g. software vulnerability of the system.

\subsection{Immutability}

The immutability property of blockchains, although standing as one of the main beneficial features, may also be abused for malicious purposes. Well-known blockchains such as Bitcoin Satoshi Vision (BSV)\footnote{\url{https://bitcoinsv.io/}} and Bitcoin \sloppy{Blockchain} have suffered from illegal data storage than cannot be deleted \cite{bbcchild,matzutt2018quantitative}. The lack of verifiable deletion mechanisms also enables DFS systems such as IPFS and IndImm\footnote{https://en.cryptonomist.ch/2019/07/29/indimm-ripple-blockchain/} to host and disseminate illegal content \cite{ijisipfs}. Therefore, neither contents nor domain names are subject to a take-down mechanism. Moreover, strategies as blacklisting domains are unpractical if the number of domains is high enough.

From a legal perspective, the GDPR does not consider the immutable nature of blockchains and DFS. In this sense, novel decentralised technologies implement features that are not aligned with current regulations and their requirements, which prevents the possibility to apply requests such as the right to be forgotten \cite{politou2018forgetting,politoublock}. Thus, the aforementioned facts make the combination of blockchain DNS and DFS systems a fertile playground for malicious practices. For instance, at the moment of writing, Emercoin supports I2P links; well-known for their anonymity, however, given the continuous rise of IPFS and other DFS solutions, blockchain DNS systems may support IPFS in the near future. The support of a permanent and distributed storage, like IPFS, with blockchain DNS, would actually make a permanent link that cannot be taken down. It is evident that the combination of both would be ideal for the distribution of illegal content as the content would become permanently available for everyone who has access to the link. It should be noted that there are already initiatives that are making this bridge available, not for illegal purposes, e.g. Unstoppable Domains\footnote{\url{https://unstoppabledomains.com}}.

\section{Analysis of real-world data}
\label{sec:experiments}
To assess the extent of these threats, an analysis of real-world data was conducted. In the first set, we used the BDNS extension\footnote{\url{https://blockchain-dns.info}} and in the second one we used the Namecoin\footnote{https://www.namecoin.org/} and the Emercoin\footnote{\url{https://emercoin.com/}} blockchain platforms. We argue that the most critical domain names are the top ones of the Alexa list, as they handle most user traffic. Therefore, if an adversary would like to take over a domain, a domain in the Alexa top 1,000 domains would provide her with the biggest impact. In what follows, we will refer to \textit{A1K} as the dataset of the Alexa top 1000 domains at the time of writing.

\subsection{Using BDNS}
BDNS is an open-source extension for Chrome and Firefox. The goal of the extension is to resolve .bit, .lib, .emc, .coin, .bazar and OpenNIC domains\footnote{\url{https://www.opennic.org/}}. The extension monitors the requests of the browser for domains. If the domain falls within the supported TLDs, then it uses a RESTful interface to resolve the IP.

Based on this concept, we created a crawler which queries uses this REST interface and tries to resolve A1K domains with any of these TLDs. The result of this search was that 464 domains out of the potential 25,000 web pages (i.e. generated from the combination of A1K with the different TLDs) were registered. These 464 web pages were mapped to 465 IPs, as one of the DNS records mapped a domain to two IPs. Interestingly, 21 of these IPs were private and 444 public ones. The latter were actually 55 unique addresses, one of which was used to resolve 220 of these web pages, and 81 belong to another IP address, signifying a high concentration. In terms of countries, these domains resolve to 15 countries, as illustrated in Table \ref{fig:dist_by_country}.

Going a step further, we browsed each of the domains. From the 464 domains, 163 did not resolve anywhere or returned an error in the server-side and 9 to a default welcome landing page of a service, e.g. IIS Server.
Then, 80 pages redirected the user to a porn web page (\url{https://iusr.co}) which belonged to the same IP address (192.243.100.192). Note that the latter IP served only this web page with the exception of one page that was down. Then, many of the pages resolved to placeholder pages. Three of them resolved to the same IP (161.97.219.84) pointing to ``Computer Rehab domain hosting'', 11 pointed to a parking domain of \url{dotbit.me} with the same IP (144.76.12.6). 67 domains were registered as part of the project New Nations \url{http://www.new-nations.net} from a single IP (178.254.31.11). The latter IP also resolved 76 more web pages that were divided into three placeholder web pages (\url{ww1.partenka.net},\url{ww17.cikidot.com}, \url{ww38.partenka.net}) with 63 in the first one, 3 in the second and 9 in the last one. Notably, from the domains that resolved to the same one listed in A1K (34), almost half of them (16) belonged to porn web sites. The rest 18 of them belonged to 11 web pages, including Wikipedia, Instagram and mega.

\begin{table}[th]
\centering
\scriptsize
\begin{tabular}{lr|lr|lr}
\toprule
\textbf{Country} & \textbf{IPs} & \textbf{Country} & \textbf{IPs}& \textbf{Country} & \textbf{IPs}\\ \midrule
DE&    238    &CA&    5&    AT    &1\\
US&    146    &SG&    3&    HK    &1\\
CN&    20    &GB&    2&    IT    &1\\
FR&    12    &NL&    2&    SE    &1\\
RU&    9    &SC&    2&    TW &    1\\ \bottomrule
\end{tabular}
\caption{Distribution by country}
\label{fig:dist_by_country}
\end{table}

\subsection{The Namecoin data}
\label{sec:namecoin}
\begin{figure}[th]
    \centering
    \includegraphics[width=0.8\textwidth]{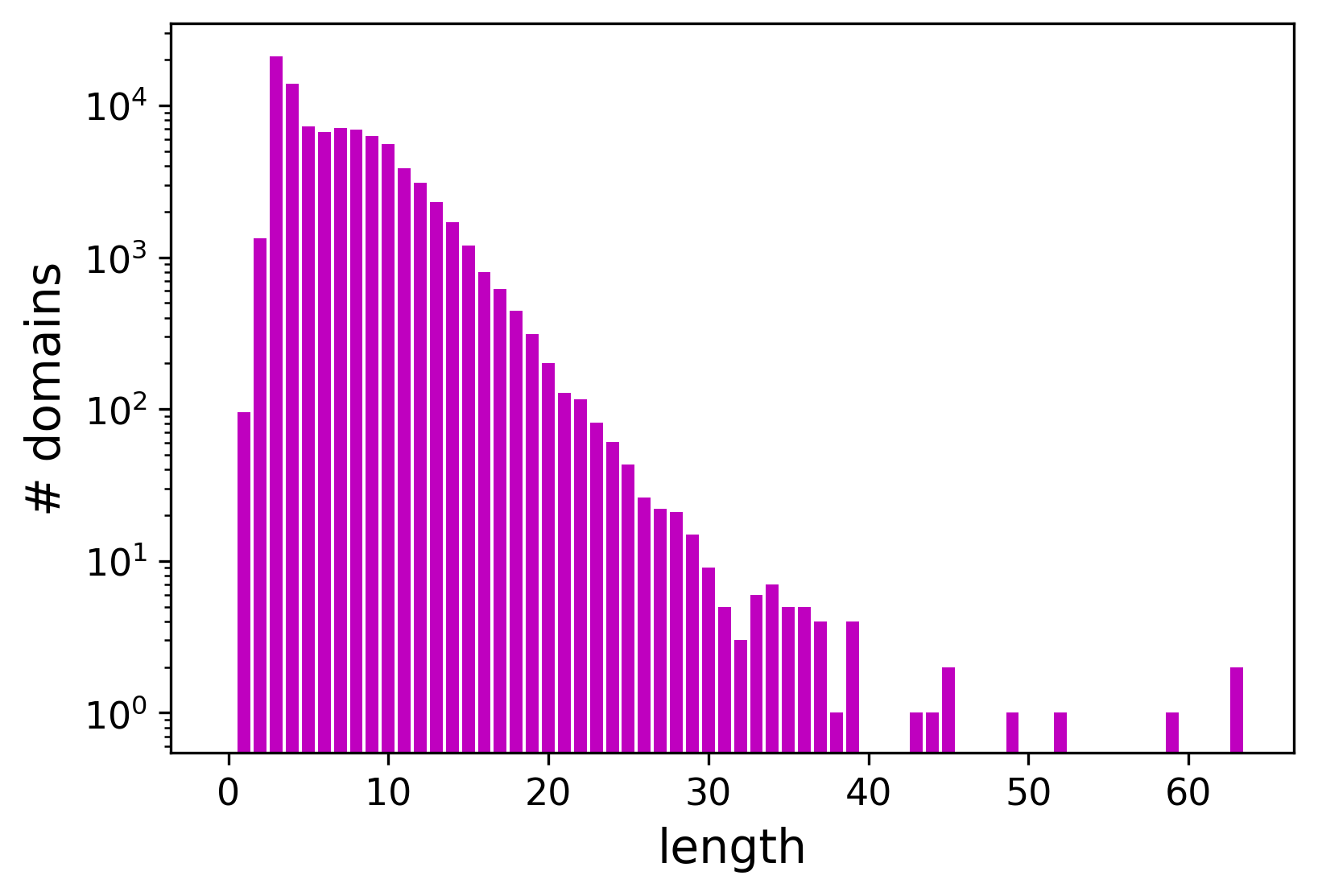}
    \caption{Length distribution of domain names registered in Namecoin. Note that values are represented in logarithmic scale.}
    \label{fig:namecoinlength}
\end{figure}

Namecoin was the first widely-used Blockchain DNS, becoming a reference point for more recent approaches such as Emercoin and Blockstack. This blockchain manages the registrar of the \texttt{.bit} TLD by means of a straightforward procedure, in which users specify the SLD that they want to register (which will be later appended with a \texttt{.bit}), as well as the resolving IP and other secondary parameters. At the time of writing this article, Namecoin has a total of 91,106 active domains (i.e. they have been recently created or periodically renewed by their owners). Nevertheless, despite the restrictions imposed by the registrar procedure and the data structure template to be added in the blockchain as well as the deviant behaviour of some users, we found some relevant statistics that showcase the potential of Namecoin as a platform to impulse illicit activities.
As a foreseeable tendency, most users opted for registering domains of low length (from the set of domains offered by ICANN, practically all SLDs with length lower than six are already registered or reserved), as described in Figure \ref{fig:namecoinlength}. Note that, as stated in Section \ref{sec:threatmalware}, this hinders procedures such as  AGDs detection. Clearly, the fact that a domain has to be renewed every certain time at a small cost, a feature which is not implemented in Emercoin as we will discuss in Section \ref{sec:emercoindata}, prevents the ownership of domains for long periods of time if there is no revenue. Nevertheless, this does not seem a constraint for some users, as seen in Table \ref{tab:addressesanddomainsnamecoin}. More concretely, a total of 86,940 addresses registered at least one domain, yet there are users that own more than 1,000 domains, which often contain the words \textit{sex}, \textit{porn}, \textit{stream}, \textit{hack} as well as other SLDs from well-known brands and companies. Although most of them do not resolve to an IP, this may change in almost real-time with a simple update.

\begin{table}[th]
\centering
\scriptsize
\begin{tabular}{lr}
\toprule
\textbf{Emercoin address} & \textbf{\# domains}\\\midrule
MyZTAGS74akZBiqYPKuvD3zGCfL8tGmXpz & 3754\\
MwyGuUCawVzCcCSoNJpWjN1Kcioq7TNM92 & 2213 \\
N256bGgH4E84P8fcEcLs4m1YCXYZb6nzAm & 1690 \\
MwkmRsY2kVjgXp2x4j9LY9fwtvXdMSaLjj & 64 \\
NJ6HHqGu9mmW25XgyGoj7V6hPoCSkQLnQ6 & 64  \\
N8sV3CJsQo83GRKw5qyBECCiFwvp1XQ2Nu & 16 \\
NBHBXLtbRLFqHRmgTufzL1aZ3ztqHyLmXH & 16 \\
MzB1bm2QDmqpmAKeaRPev4QxAxTWj1kZRi & 14 \\
NEm1R3yMmvGckDCa9Tt9XY8Yor3tEVfReb & 14 \\
NKesEinH7phBMdu3XT5KTw5eQtRoXnxhT6 & 10 \\\bottomrule
\end{tabular}
\caption{Top 10 addresses in Namecoin with most registered domains.}
\label{tab:addressesanddomainsnamecoin}
\end{table}

Due to the fact that OpenNIC DNS servers, as well as a set of browser extensions, resolve \texttt{.bit} domains, we found some other relevant statistics that need to be discussed to give a clear picture of what is happening behind the curtains. Since these findings include data extracted from Emercoin, we will address them in the following section.

\subsection{The Emercoin data}
\label{sec:emercoindata}
The Emercoin blockchain is one of the most well-known services for domain registration. In total, the blockchain contains 53,408 records at the time of writing. Interestingly, although the naming requirements of Emercoin specify that only lowercase alphanumeric ASCII characters are allowed, the chain contains case sensitive domains not only for the advertised TLDs but for standard TLDs like \texttt{.com}. The distribution of the domains is illustrated in Table \ref{tab:lexicaldomains}. In this regard, we observed that most of the addresses registered one or two domains (i.e. a total of 43,543 addresses registered at least one domain in Emercoin), some addresses registered more than 1,000 domains, as showed in Table \ref{tab:addressesanddomains}. Many of these records contained an IP, an email address, or a note advertising that the domain is for sale. More concretely, by querying the Emercoin blockchain, we found that up to 567 domains contain the words ``for sale'' in their \textit{value} field, and in most cases an email to contact. Moreover, when searching for ``\$" in the \textit{value} field, the search returned more than 100 domains with a specific sale value. Finally, correlating the A1K dataset with the Emercoin chain returned 1,045 domains, which correspond to 328 unique SLDs registered with different TLD variants.

\begin{table}[th]
\centering
\scriptsize
\begin{tabular}{lr}
\toprule
\textbf{Feature} & \textbf{Registered domains} \\\midrule
\texttt{.com} TLD & 44\\
Punycode (xn--)  & 1261 \\
Capital letters & 316 \\
Whitespace character & 35 \\\bottomrule
\end{tabular}
\caption{Lexical statistics for domain names registered in Emercoin.}
\label{tab:lexicaldomains}
\end{table}

\begin{table}[th]
\centering
\scriptsize
\begin{tabular}{lr}
\toprule
\textbf{Emercoin address} & \textbf{\# domains}\\\midrule
ETkxi1X1CeX2QDSWp3CDmuDj7jJZtftfNF & 3565\\
EKzDF4RAHat8tWdQGbvR9zm7PJrHcth7Rm & 2688 \\
EQADxQhroZwGnQAyirFtNbwwjoykciFqv3 & 253 \\
EYBExDLR3aqZunRj6NuyRC9TXt8NHKKXWZ & 196 \\
ENnpjY8YQr5rvKNc1TY6kkBwsDZXwmEiY2 & 150  \\
EWwX61CW9TorzZ7Dy1dmnfKYPxz7dBMGxJ & 137 \\
EaQkdxCMPVzMXtTFqYaQxV7wQ1qqLy8aXF & 58 \\
ELRNsgvTbV83MyPdD5ACf1xyemLFV7Sued & 53 \\
ESCWovPDaX55KCpX3bdkKWqbH4zBEiwNrd & 47 \\
EZKCa2ELZpPoNPFQrsHQXszFLrPEf9Q5vJ & 44 \\\bottomrule
\end{tabular}
\caption{Top 10 addresses in Emercoin with most registered domains.}
\label{tab:addressesanddomains}
\end{table}

\begin{figure}[th]
    \centering
    \includegraphics[width=0.8\textwidth]{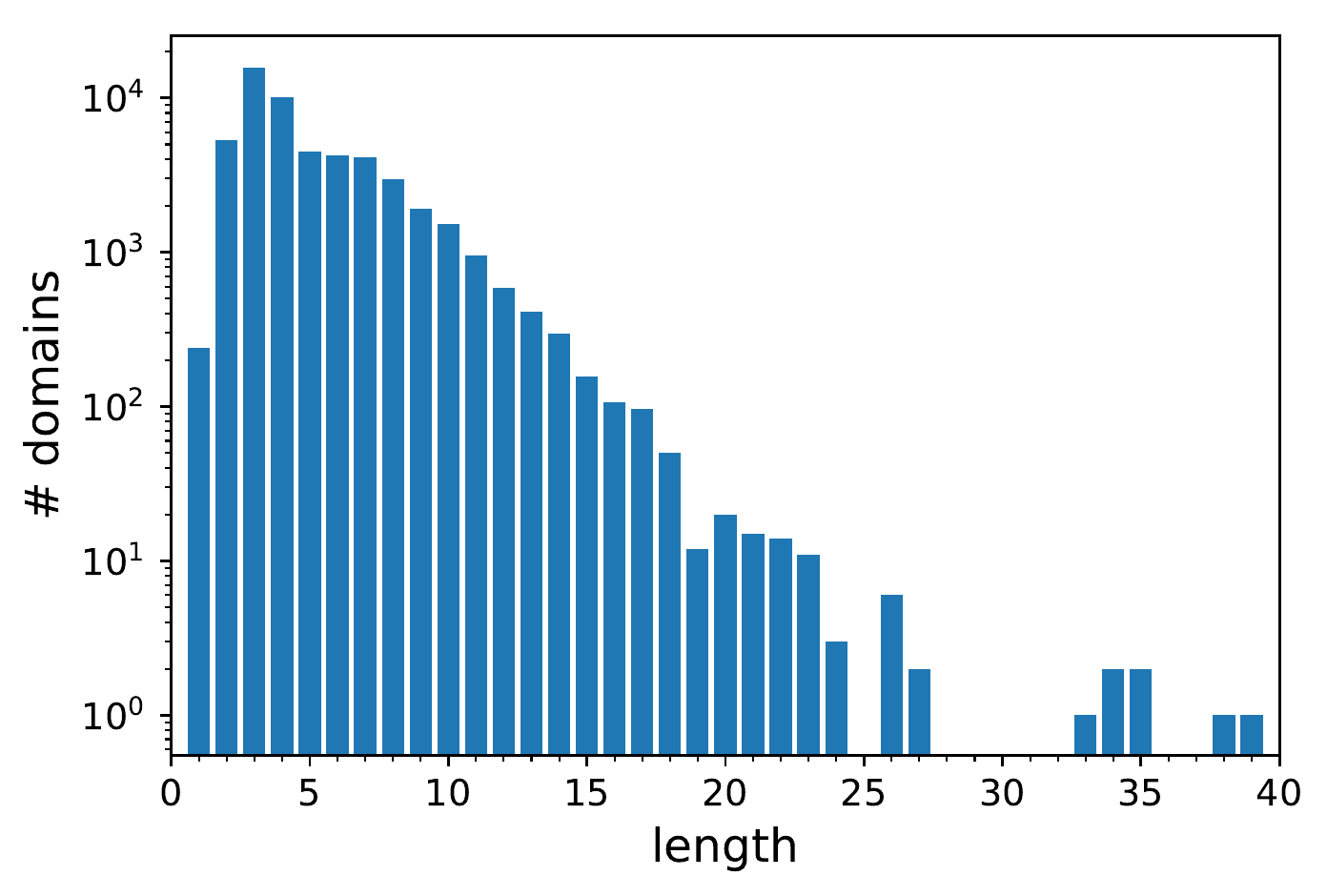}
    \caption{Length distribution of domain names registered in Emercoin. Note that values are represented in logarithmic scale.}
    \label{fig:emerlength}
\end{figure}

The domain name length distribution is depicted in Figure \ref{fig:emerlength}. Notably, most of the domains have lengths below five, with three letters being the most registered domains (as in the case of Namecoin). As previously stated, these SLD are no longer available in ICANN, since they are already registered, and are among the first to be registered once a new TLD appears. Given the high correlation with ICANN domains, it is expected that many of them, if they do not belong to the corresponding ICANN owners, are highly likely to be used for malicious activities such as phishing or cybersquatting.

\begin{figure}[th]
    \centering
    \includegraphics[width=\textwidth]{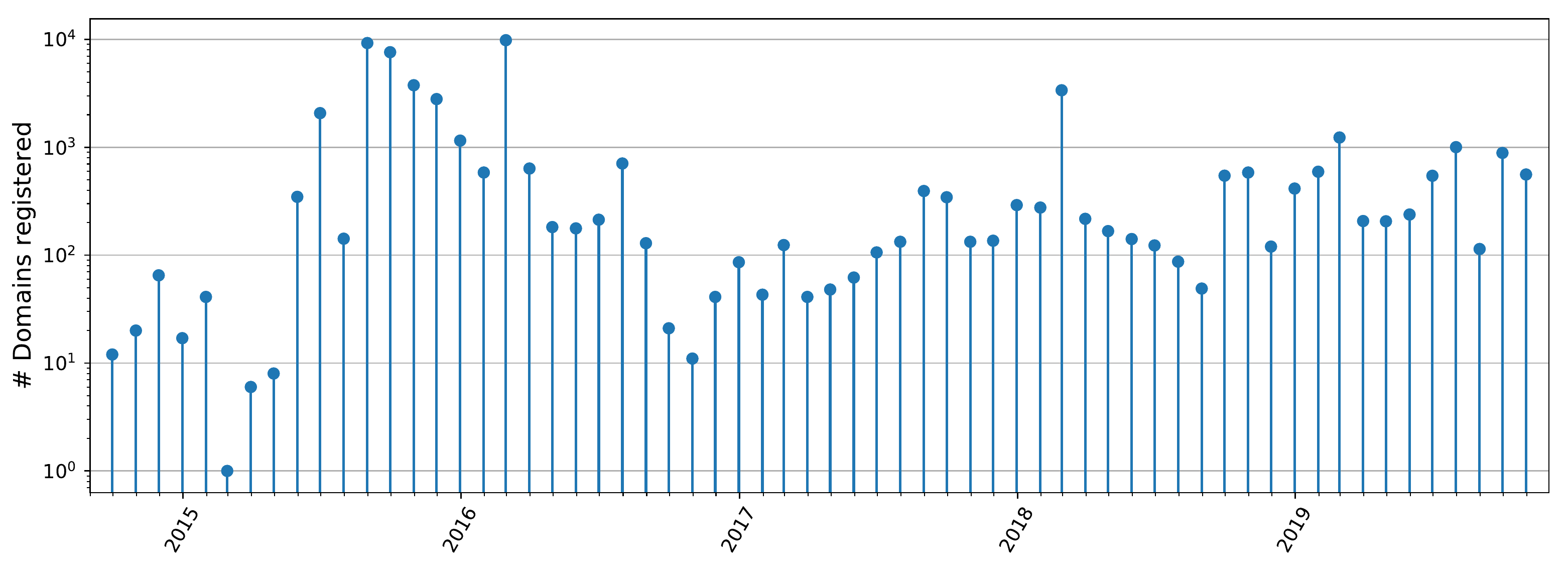}
    \caption{Timeline of registered domains in Emercoin. Note that values are represented in logarithmic scale.}
    \label{fig:timelineemer}
\end{figure}

Finally, some statistics of the domain registering behaviour over time are depicted in Figure \ref{fig:timelineemer}, which shows the domains registered since the beginning of the blockchain until October 2019. Notably, we can see some peaks in its lifetime. 

\begin{figure}[th]
    \centering
    \includegraphics[width=\textwidth]{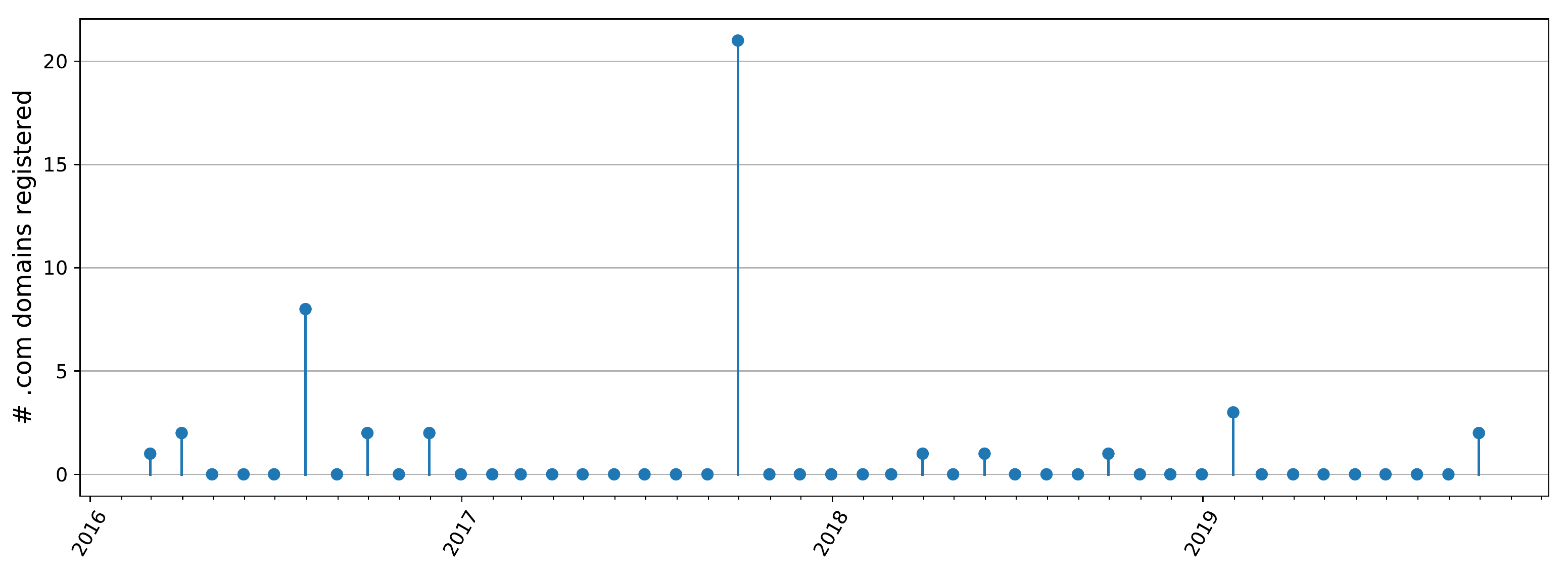}
    \caption{Timeline of \texttt{.com} domains registered in Emercoin.}
    \label{fig:timelineemercom}
\end{figure}

The distribution in time of the domains registered with \texttt{.com} was also investigated. As seen in Figure \ref{fig:timelineemercom}, such practices, although not alarmingly numerous, are still active in 2019. Therefore, the registrar system still allows the registration of domains with TLDs different than these offered by Emercoin. This problem is related to several of the threats discussed in Section \ref{sec:threats}, such as the vulnerabilities with the underlying registrar, which may enable further malware and phishing campaigns, as well as cybersquatting.

Finally, we computed some global statistics for Namecoin and Emercoin. Currently, there are more than 140K domain names registered in both blockchains, but only 5,266 have an IP associated with them\footnote{https://blockchain-dns.info/explorer/}. Out of these 5,266, we computed the distribution of TLDs and depicted the results in Table \ref{tab:dist_by_tld}. We can observe that most of domains belong to \texttt{.coin}, \texttt{.bit}, \texttt{.lib}, \texttt{.bazar}, and \texttt{.emc}. Note that some of the other TLDs should not be ``available", with special regard of the whitespace character. Next, we wanted to explore what the distribution of IPs controlling these domains is. In this regard, the top 15 IPs used for that purpose are described in Table \ref{tab:dist_by_ip}. Notably, we may observe that \textit{192.243.100.192} is the IP to which most domains resolve (i.e. a total of 1957 domains).

\begin{table}[th]
\centering
\scriptsize
 \setlength{\tabcolsep}{5pt}
\begin{tabular}{lr|lr|lr}
\toprule
\textbf{TLD} & \textbf{Number} & \textbf{TLD} & \textbf{Number} & \textbf{TLD} & \textbf{Number}\\ \midrule
coin       &                1261 & \$              &1  & net                               &1 \\
bit        &                1045 &oz               &1    &ln                         &1\\
lib       &               1017 & \*                 &1   & in          &1    \\
bazar      &               998 &bbs                &1  &9988                             &1 \\
emc        &              861 &news                &1   &kib                             &1 \\
i2p       &              19 &ua                   &1  &fashion                        &1 \\
neo       &             14 &luxsocks              &1 &woshiwo321                    &1 \\
com        &             8 &mayun                  &1  &name                         &1 \\
onion     &            3 &years                   &1 & www         &   1  \\
cn       &           3 &pi                       &1  & cion        &       1 \\
coin      &          2 &aaatttaaa                 &1 & mec       &      1 \\
eth           &         2 &io                         &1  & su           &     1\\
enc          &        1 &liib                        &1  & biz            &    1  \\
org          &1 & linux                               &1  & 1010         &   1 \\ \bottomrule
\end{tabular}
\caption{Distribution of TLDs resolving to an IP in both Emercoin and Namecoin.}
\label{tab:dist_by_tld}
\end{table}

\begin{table}[th]
\centering
\scriptsize
\begin{tabular}{lr|lr}
\toprule
\textbf{IPs} & \textbf{Domains} &\textbf{IPs} & \textbf{Domains} \\ \midrule
192.243.100.192 & 1957 & 78.107.255.15 & 53\\
144.76.12.6 & 448 & 192.241.241.153 & 45\\
202.108.22.5 & 402 & 202.108.8.82 & 45\\
192.227.233.13 & 340 & 81.2.247.158 & 45\\
178.128.220.134 & 144 & 94.242.60.7 & 37\\
185.31.209.8 & 88 & 185.61.138.167 & 32 \\
178.32.148.152 & 67 & 46.29.251.130 &29\\
92.63.101.1 & 53 &     & \\\bottomrule
\end{tabular}
\caption{Top 15 IPs to which domains resolve in both Emercoin and Namecoin.}
\label{tab:dist_by_ip}
\end{table}

\section{Discussion and Countermeasures}
\label{sec:conclusions}

Arguably, the aforementioned threats seem to portray an obscure future. In what follows, we propose a set of mitigation strategies and mechanisms for each of the identified threats.

As described in Section \ref{sec:emercoindata} the Emercoin registrar allows some theoretically forbidden patterns and characters, including the \texttt{.com} TLDs. These practices, although uncommon, are still active, as seen in Figure \ref{fig:timelineemercom}. In the case of Namecoin, the periodic renewal mechanism, as well as the fact of only controlling one TLD, enables higher control of data, yet both blockchains have similar patterns and user behaviours as analysed in Sections \ref{sec:namecoin} and \ref{sec:emercoindata}. As such, more robust mechanisms have to be implemented in the future to avoid deviant behaviours. These mechanisms should cover the whole registrar procedure in an end-to-end manner, from the auction systems (e.g. with robust smart contracts and revocation mechanisms, triggered according to a majority) to the proper checking of the data structures stored in the blockchain so that malicious/unexpected information cannot be inserted. Other solutions and functionalities such as forks, which will be later described for the case of the immutability threat, could also be adopted.  

In the case of cybersquatting, several strategies have been implemented by systems like Handshake, in which they pre-reserved the top 100k Alexa domains. Other similar policies may be implemented in future decentralised DNS systems as well as a controlled flow of domains being registered, to prevent users from registering arbitrary amounts of domains. Due to the unrestricted nature of Blockchain DNS systems, users may register the most used SLDs and append one of the multiple TLDs offered by the new blockchain DNS registrars. As previously stated, the appearance of blockchain DNS systems which aim to register and resolve all the domain spectrum (both in terms of SLDs and TLDs), may create different versions of the Internet. In this scenario, the challenge of controlling the domain name registration as well as the resolution will require unprecedented security and privacy mechanisms.

Email had always accommodated a noteworthy attack surface due to the lack of security considerations since its inception. The evolution of email security at some point called upon the DNS infrastructure to address integrity and authentication issues as an attempt to prevent certain types of spam and phishing. Email security policies and protocols such as the Sender Policy Framework (SPF), Domain Keys Identified Mail (DKIM) and Domain Message Authentication Reporting (DMARC) which depend upon DNS can be extended and adapted to force checks on domains and prevent domain spoofing attempts. In addition, the email clients should include scanning and checking functionality to distinguish between the different emerging \textit{parallel} Internets attributed to different blockchain DNS entries. The email servers (and MTAs in general) could enforce tighter policies by requiring properly configured DMARC services. In essence, the email ecosystem could act in this instance as the gatekeeper prior to entering the blockchain DNS controlled realm.

The decentralised nature of blockchain DNS is expected to change and improve the botnets' C2 communication channels, by providing more effective rendez-vous algorithms than the current DGAs. Fewer NXDomain responses, covert channels and encrypted communications are expected. Traffic analysis similar to the one described by \cite{cose2019} is expected to be less effective. This new state of play would require more proactive approaches such as hunting for synthesised IoC type of patterns in the blockchain itself, not only limited in the domain information, but also all available metadata. The immutability of the blockchain would allow to continuously and reliably study the botnets' modus operandi and respond with mitigation actions.

The immutability of blockchains requires other approaches to counter malicious records. Although less popular, forks are a well-known mechanism to ``delete'' data from the blockchain \cite{politoublock}. Nevertheless, forks are used only in exceptional cases, and are not considered to be an efficient solution, since they add a prohibitive overhead to the system, especially if the number of deletion requests is high. Other strategies regarding the block consolidation mechanism (the number of blocks created in front of the actual block for it to be considered safe) can also be explored, yet, again, they could hinder the efficiency of the system. In terms of blockchains, technical efforts to circumvent immutability while preserving their inherent security are steadily emerging \cite{politoublock}.


Finally, it should be emphasised that in order for such initiatives to become mainstream and not a tool for cybercrime, they need to build trust on their services. At their current form it is evident that both Namecoin and Emercoin have already many issues as their users face many privacy and security issues. Therefore, moderation solutions must be deployed to protect the name of the emerging ecosystem. The moderation may prevent poisoning of the chains and removal of malicious records making the users trust the provided services.

\section{Conclusion}

When a disruptive technology such as blockchain enters the realms of one of the core Internet services such as DNS, it is imperative that the security community invests a significant amount of effort to study and investigate the security implications. The DNS hijacking incident back in 2014 where 300K routers were compromised\footnote{\url{https://www.theregister.co.uk/2014/03/04/team_cymru_ids_300000_compromised_soho_gateways/}}, albeit having a high impact to businesses, is minuscule compared to the potential damage malicious actors can cause when the blockchain DNS becomes widely accepted.

This paper attempted to tessellate the emerging threats and provide insight into the associated risks introduced by moving from a centralised to a fully decentralised DNS. From a forensic investigation perspective, the use of blockchain is a mixed blessing; on the one hand, some of the evidence will be stored in a forensically sound manner. On the other, the introduction of yet another technology into the Internet backbone will not only increase the complexity leading to a potentially greater attack surface but will also result in significant attribution challenges.

\section*{Acknowledgements}
This work was supported by the European Commission under the Horizon 2020 Programme (H2020), as part of the projects CyberSec4Europe (\url{https://www.cybersec4europe.eu}) (Grant Agreement no. 830929), \textit{LOCARD} (\url{https://locard.eu}) (Grant Agreement no. 832735) and \textit{ECHO}, (\url{https://echonetwork.eu}) (Grant Agreement no 830943).

The content of this article does not reflect the official opinion of the European Union. Responsibility for the information and views expressed therein lies entirely with the authors.

\end{document}